# Improving Seek Time for Column Store Using MMH Algorithm

Tejaswini Apte

Sinhgad Institute of Business
Administration and Research
Kondhwa(BK),Pune-411048

Dr. Maya Ingle

Devi Ahilya VishwaVidyalay
Indore

Dr. A.K.Goyal

Devi Ahilya VishwaVidyalay
Indore

*Abstract*—Hash based search has, proven excellence on large data warehouses stored in column store. Data distribution has significant impact on hash based search. To reduce impact of data distribution, we have proposed Memory Managed Hash (MMH) algorithm that uses shift XOR group for Queries and Transactions in column store. Our experiments show that MMH improves read and write throughput by 22% for TPC-H distribution.

*Keywords- Load; Selectivity; Seek; TPC-H; Algorithms; Hash.*

## I. INTRODUCTION

Searching in Column Store (CS) is greatly influenced by the address lookup process. Hashing algorithms have been widely adopted to provide fast address look-up process [2, 3, 8]. Bob Jenkins' hashing algorithm processes the key twelve octets at a time; the post processing step is slightly more complex because of handling of partial final block [14] in CS. However, it is possible to improve the throughput rate for fast address lookup in CS.

For various data warehouse applications, address lookup performs major role in performance measurement. The related and existing techniques of hashing and lookup are discussed in Section 2. Hash scan participates in performance of CS; Section 3 summarizes the hash scan for simple and complex queries. The proposed algorithm is an improved version of Jenkins' algorithm named as MMH. The informal and formal description of algorithm is discussed in Section 4. Case study was presented to show the effectiveness of our algorithm MMH with the help of implementation details in Section 5. Result analysis of MMH over Jenkins' is discussed in Section 6. Finally, we conclude with future work in Section 7.

### I.RELATED WORK

Hashing has been used most successfully to avoid block conflicts in interleaved parallel memory systems used in multiprocessors and vector processors. Linear skewing functions, computes the block number using integer arithmetic [2, 3]. Stride patterns are mapped conflict-free when the stride and the number of memory blocks are relative primes [4].

To minimize the latency in computing per-block address, fragmentation was introduced in the Burroughs Scientific Processor, however it wastes 1/17th of the memory [5]. Fragmentation and complex block number computations are not necessary to obtain conflict free access to stride patterns. It has been observed that some particular types of XOR-based hash functions that are based on the division of binary polynomials, can simultaneously map a large number of stride-based patterns conflict-free [6]. XOR-based interleaving functions mainly focused on constructing a conflict-free hash function for several patterns complete with success [15, 8]. Bob Jenkins' hash produces uniformly distributed values for the hash tables [14]. However, literature reveals that there is a scope to improve the seek time of Jenkins algorithm for Column Store.

## II. COLUMN STORE HASH SCAN

This section describes Hash Scan for simple and complex queries both for column store.

### A. Hash Scan for Simple Queries

The complexity of hash scan is highly influenced by the size of data warehouse. Hash function may use partial or entire record as key to generate hash value. The parameters for hash based search are selectivity and cardinality for the given query. For shift XOR, with uniform distribution, if the key is having n values, probability density function (pdf) is:

$$\text{Selectivity} = n/ \text{ (number of distinct values)}$$

$$\text{Pdf (n)} = 1/ \text{ selectivity}$$

### B. Hash Scan for Complex Queries

Assume the given relation has multiple attributes stored in CS architecture. Let AK is the length of attribute, LID is the length of the tuple identifier or primary key and MROW is the matched row of second segment.

The number of seeks for given query is expressed as:

- Number of seeks required to retrieve tuples from the scanned segment.

  Number of seeks = ((numberOfRows) *AK + LID)* blocksize
- Number of seeks required to retrieve the remainder of the original tuples for those transactions which require it.

Number of seeks = ((numberOfRows) *AK +        LID)* blocksize + ((MROW)*AK+LID)*blocksize

## III. PROPOSED ALGORITHM - MMH

MMH algorithm is designed and tested with varying selectivity and cardinality of TPC-H distribution. The performance improvements could be demonstrated by executing following query on TPC-H schema with & without MMH algorithm.

```
select
            s_acctbal,
            s_name,
            n_name,
            p_partkey,
            p_mfgr,
            s_address,
            s_phone,
            s_comment
from
            part,
            supplier,
            partsupp,
            nation,
            region
where
            p_partkey = ps_partkey
            and s_suppkey = ps_suppkey
            and p_size = 29
            and p_type like '% BURNISHED TIN'
            and s_nationkey = n_nationkey
            and n_regionkey = r_regionkey
            and r_name = 'MIDDLE EAST'
            and ps_supplycost = (
                        select
                                    min(ps_supplycost)
                        from
                                    partsupp,
                                    supplier,
                                    nation,
                                    region
                        where
                                    p_partkey = ps_partkey
                                    and s_suppkey = ps_suppkey
                                    and s_nationkey = n_nationkey
                                    and n_regionkey = r_regionkey
                                    and r_name = 'MIDDLE EAST'
            )
order by
            s_acctbal desc,
            n_name,
            s_name,
            p_partkey;
```

### A. Informal Description

The proposed algorithm MMH is broadly designed with four functions:

- query(TPC-H-Q) Input parameter is a TPC-H query and return a valid sql query as output. This function is necessary to provide query for generation of hash value to improve search time.

- strHash(q) Input parameter is a valid sql query, this function uses CSXOR function to change the query to appropriate hash value. Primitive operations on database points to BUN heap, contains the atomic values inside the two columns. Fixed-sized atoms, reside directly in the BUN heap.

- HEAPalloc(d, size, 1) Input parameters are the memory heap and size. This function carries out checks for allocation of memory.

- CSXOR (h,s) Input parameters are memory heap and query. Execution generates hash value and is placed in passed heap.

### B. Formal Description - MMH

/* Memory Managed Hash (*MMH*) Algorithm stores hash values in memory location for TPC-H schema query processing */
/* **Main program begins** */
**main()**
{
  s=query(TPC-H);
  **strHash (char \*s)**
  {
      /* Declaration of variables */
      Heap d, size = 1<<10 * sizeof(stridx);
      /* Checking  allocation size */
      if  HEAPalloc(d, size, 1) >= 0)
      {
          d->free = 1<<12 sizeof(stridx);
          /* Declare and initialize Binary UNits     (BUN)*/
          BUN res=1<<10-1;
          /* Call a function with string *s* and store in BUN */
          res=CSXOR(*d*,s);
          /* stores hash values  in heap d with its base
          value, base position and free space*/
          memset(res->base, 0, res->free);
          }
  }
}
**end.**
/* **End of main** */
*CSXOR*(Heap *h*, const  char *v*)
/* This function performs string search with shift XOR operation; with input parameters h as memory space and v as constant string; It outputs generated hash values */
begin
{
  /* Declaration of variables */
  stridx_t \*ref, \*next;
  /*Extend memory allocation by allocating more binary units BUN */
  EXLEN=BUN size+1<<3-1;
  size_t exlen=*h*->hashash?EXLEN:0;
  /* Initializing binary units*/
  BUN off;
  off=1<<10-1;
  /* Shift XOR operation for generating hash values and to search
          string */
  hkeyvalues=0;
  for i = h to v do
  {
      hkeyvalues \^= (hkeyvalues << 10) +
                  (hkeyvalues >> 7) +v;
      h.base=hkeyvalues;
  }
```





```
/* Searching for string in heap */
for ref = h.base+off to h.length do
{
  ref=next;
  next=(h.base+ref);
  if STRCMP(v, (str) (next + 1) + exlen) == 0)
  return  ((sizeof(stridx_t) + *ref + exlen) >>3);
  }
 }
end.
```

## IV. CASE STUDY

MMH is designed from the shift-XOR class of hashing function. To support the hypothesis, we experimentally evaluate the MMH on real data sets i.e. TPC-H schema. In our experiments, we have focused on certain table sizes and load factors, to allow comparisons with original algorithm. We first investigated average search lengths for successful and unsuccessful search. The MMH results are compared to Jenkins' algorithm (Table 1 and Table 2). As can be seen, proposed algorithm performs better for TPC-H schema.

TABLE I.        RESPONSE OF EXECUTION OF QUERIES

| TPC-H Queries | Jenkins' time (in ms) | MMH time (in ms) |
|---|---|---|
| 2 | 196.959 | 92.73 |
| 3 | 1100 | 900 |
| 4 | 611.088 | 541.846 |
| 5 | 929.93 | 900 |
| 6 | 601.881 | 540.692 |
| 7 | 1000 | 900 |
| 8 | 227.257 | 196.401 |
| 9 | 855.851 | 672.403 |
| 10 | 4200 | 3000 |
| 12 | 850.772 | 727.156 |
| 16 | 423.415 | 305.074 |
| 17 | 396.955 | 258.854 |
| 19 | 3400 | 2500 |
| 21 | 1000 | 772 |

TABLE II.        COMPARISON OF TRANSACTION LOAD TIME

| Relation | Jenkins' (in ms) | MMH (in ms) |
|---|---|---|
| Region | 179.696 | 140 |
| Nation | 136.036 | 102.321 |
| Supplier | 272.617 | 202.886 |
| Customer | 1600 | 1000 |
| Part | 1600 | 1200 |
| PartSupp | 25300 | 18200 |
| Orders | 30000 | 17700 |
| LineItem | 140000 | 100000 |

## II. RESULT ANALYSIS

The proposed algorithm performs uniformly and efficiently independently of data size. From experiments with large sets of keys we have observed that with poorly chosen hashing function, performance can deteriorate markedly as the number of keys increases (Figure 1). Experimental results for the expected length of the load search time (LST) values vary significantly between runs. We chose a random set of TPC-H schema keys, the distribution of LST values is even narrower.

MMH improvement to average LST is 30% on Red Hat Linux 2.4 GHz Intel processor and 1GB of RAM. (Figure 2). To our knowledge these are the first experiments testing these predictions.

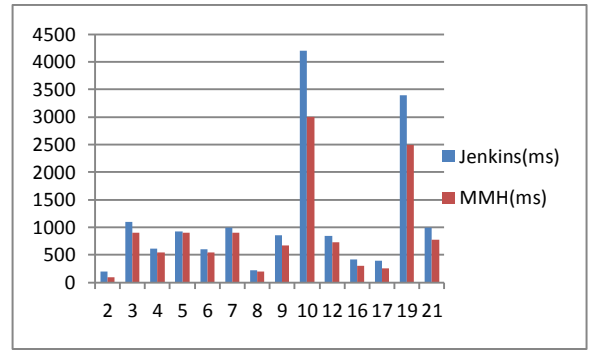

Figure 1.    Result Analysis for Transaction Query Time

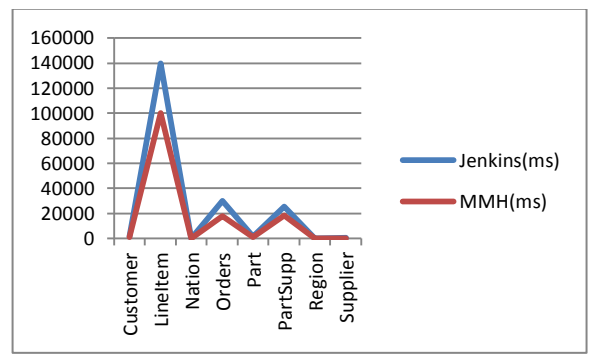

Figure 2.    Figure 2: Result Analysis for Transaction Load Time

## V. CONCLUSION AND FUTURE WORK

The proposed algorithm is a generic search algorithm for CS data storage. The algorithm is designed specifically for use in query intensive environment. A key design principle of MMH to improve the throughput by minimizing the disk seeks. To achieve we used the hash function of shift-XOR class. We experimentally demonstrated gain in performance by MMH. The continued evolution of hard disk technology should make such performance advantages clearer in the future. The most obvious avenue for future work is an extension of MMH algorithm for multiple instances of CS. The most significant question that must be addressed when extending the MMH to a multi-instance environment is handling synchronization for various disks seeks.


REFERENCES

[1] H. Vandierendonck and K. De Bosschere, "Randomized Caches for Power-Efficiency," IEICE Trans. Electronics, vol. E86-C, no. 10, pp. 2137-2144, 2003.

[2] D.J. Kuck, "ILLIAC IV Software and Application Programming," IEEE Trans. Computers, vol. 17, no. 8, pp. 758-770, Aug. 1968.

[3] G.S. Sohi, "High-Bandwidth Interleaved Memories for Vector Processors—A Simulation Study," IEEE Trans. Computers, vol. 42, no. 1, pp. 34-44, Jan. 1993.

[4] D.J. Kuck and R.A. Stokes, "The Burroughs Scientific Processor (BSP)," IEEE Trans. Computers, vol. 31, no. 5, pp. 363-376, May 1982.

[5] D.H. Lawrie and C.R. Vora, "The Prime Memory System for Array Access," IEEE Trans. Computers, vol. 31, no. 5, pp. 435-442, May 1982.







[6] B.R. Rau, "Pseudo-Randomly Interleaved Memory," Proc. 18th Ann. Int'l Symp. Computer Architecture, pp. 74-83, May 1991.

[7] R. Raghavan and J.P. Hayes, "On Randomly Interleaved Memories," SC90: Proc. Supercomputing '90, pp. 49-58, Nov. 1990.

[8] J.M. Frailong, W. Jalby, and J. Lenfant, "XOR-Schemes: A Flexible Data Organization in Parallel Memories," Proc. 1985 Int'l Conf. Parallel Processing, pp. 276-283, Aug. 1985.

[9] N. Topham and A. Gonza´lez, "Randomized Cache Placement for Eliminating Conflicts," IEEE Trans. Computers, vol. 48, no. 2, pp. 185-192, Feb. 1999.

[10] Jenkins Bob "A hash function for hash Table lookup".

[11] A. Gonza´ lez, M. Valero, N. Topham, and J.M. Parcerisa, "Eliminating Cache Conflict Misses through XOR-Based Placement Functions," Proc. 1997 Int'l Conf. Supercomputing, pp. 76-83, July 1997.

[12] Expected worst-case performance of hash files. Copmuter journal, Volume 25,Number 3, pages 347-352,1982

[13] VY. Lum. General performance analysis of key-to address transformations method using an abstract file concept. communication of the ACM, volume 16, Number10, pages 603-612,1973